\documentclass[twocolumn,prb]{revtex4}
\usepackage{graphicx}
\newcommand{\be}{\begin{equation}}
\newcommand{\ee}{\end{equation}}
\newcommand{\bea}{\begin{eqnarray}}
\newcommand{\eea}{\end{eqnarray}}
\newcommand{\HH}{{\cal H}}

\newcommand{\p}{\partial}

\newcommand{\lb}{\left[}
\newcommand{\rb}{\right]}
\newcommand{\lp}{\left(}
\newcommand{\rp}{\right)}

\renewcommand{\phi}{\varphi}
\renewcommand{\epsilon}{\varepsilon}
\renewcommand{\vec}[1]{{\bf #1}}
\renewcommand{\cite}[1]{[\onlinecite{#1}]}

\begin{document}

\title{Atomic collapse, Lorentz boosts, Klein scattering, and other quantum-relativistic phenomena in graphene}
\author{
Andrei Shytov${ }^1$, Mark Rudner${ }^2$, Nan Gu${ }^3$, Mikhail Katsnelson,${}^4$ Leonid Levitov${ }^3$}
\affiliation{
${}^1$ Department of Physics, University of Utah, Salt Lake City, Utah 84112 \\
${}^2$ Lyman Laboratory, Physics Department, Harvard University, Cambridge MA 02138 \\
${}^3$ Department of Physics,
 Massachusetts Institute of Technology, 77 Massachusetts Ave,
 Cambridge, MA 02139 \\
${}^4$ Radboud University of Nijmegen, Toernooiveld 1
6525 ED Nijmegen, The Netherlands
}

\begin{abstract}
Electrons in graphene, behaving as massless relativistic Dirac particles, 
provide a new perspective on the relation between condensed matter and high-energy physics. We discuss atomic collapse, a novel state of superheavy atoms stripped of their discrete energy levels, which are transformed into resonant states. Charge impurities in graphene provide a convenient  condensed matter system in which this effect can be explored. Relativistic dynamics also manifests itself in another system, graphene p-n junctions. We show how the transport problem in the presence of magnetic field can be solved with the help of a Lorentz transformation, and use it to investigate magnetotransport in p-n junctions. Finally, we review recent proposal to use Fabry-P\'erot resonances in p-n-p structures as a vehicle to investigate Klein scattering, another hallmark phenomenon of relativistic dynamics. 
\end{abstract}

\maketitle

\section{Introduction}
\label{sec:intro}

The unique properties of graphene, a single atomic layer of carbon \cite{Novoselov04}, attract a lot of attention and interest from researchers from diverse fields. What makes this material special is 
its transport characteristics, such as high mobility and tunable carrier density, which offer new exciting opportunities for nanoelectronics \cite{Geim07}. At the same time, charge carriers in graphene exhibit many unusual properties, posing interesting questions of fundamental interest \cite{Beenakker07}. 

Undoped graphene is 
a semimetal in the solid state nomenclature, with conduction and valence bands joined together at the Fermi points, which are the symmetry points of the Brillouin zone known as $K$ and $K'$. Near these points, the conduction band has linear dispersion, which can be modeled by the Dirac equation. Charge carriers in graphene are thus 
described as massless relativistic fermions, 
with an effective ``speed of light'' $v \approx 10^6{\rm m/s}$, which is about $300$ times less than $c$, the speed of light in vacuum.  

This description in terms of massless Dirac particles leads to an interesting analogy
between the physics of graphene and that of high-energy relativistic particles, 
offering a possibility to observe quantum-relativistic
phenomena in condensed matter experiments. Various interesting phenomena associated with relativistic fermions, such as the half-integer quantum Hall effect \cite{Gusynin05,Peres06}, Klein tunneling \cite{Katsnelson06,Cheianov06} and scattering\cite{Lorentz,Klein_in_FP}, atomic collapse \cite{Collapse1,Collapse2,Pereira07a}, gauge fields and topological defects\cite{Morpurgo06,Cortijo07}, can be translated directly into transport properties of graphene. 

Compared to other two-dimensional electron systems, the linear dispersion in graphene, $\epsilon = \pm v |{\bf p}|$, leads to much higher quantization energies. This is the case, for example, for the quantized energies of Landau levels. The Dirac Hamiltonian
in a uniform magnetic field yields $\epsilon_n=\pm v\sqrt{2e\hbar B n}$, $n=0,1,2...$, with an interlevel spacing that can exceed $1000\,{\rm K}$ in a field of about $6$ Tesla.
As a result, quantized Hall transport in graphene can be observed at room temperature \cite{Novoselov07}.

The ``Diracness'' of charge carriers plays a fundamental role in determining the electronic properties of graphene \cite{Beenakker07}. 
In terms of the wavefunction amplitudes on the A and B sublattices of the carbon honeycomb lattice, $\psi=(\psi_A,\psi_B)$, the Hamiltonian takes the form \cite{DiVincenzo84}
%
\be\label{H_D}
H_{K(K')}=\zeta \lp \begin{array}{cc}
0 & vp_-\cr vp_+& 0
\end{array}\rp 
,\quad
p_\pm=p_1\pm ip_2
\ee
where $\zeta=+1(-1)$ for the point $K(K')$. Starting with the Schr\"odinger equation $i\hbar\p_t\psi=H\psi$, and multiplying it by the matrix $\sigma_z$, we can write it in the canonical form of a relativistic Dirac equation for a free particle
\be\label{eq:Dirac_eqn}
\gamma^\mu\p_\mu\psi=0
,\quad
x^\mu=(vt,x_1,x_2).
\ee
The $2\times2$ matrices $\gamma^\mu$ satisfy the anticommutation relations $\{\gamma^\mu,\gamma^\nu\}=g^{\mu\nu}$, where $g^{\mu\nu}$ is the metric tensor.

The analogy with special relativity extends to problems involving coupling of electrons to external electric or magnetic fields. 
Rewriting the Schr\"odinger equation $i\hbar\p_t\psi=\lb H(\vec p-e\vec a)+e\phi\rb\psi$ in the form (\ref{eq:Dirac_eqn}) yields the Dirac equation in a background electromagnetic field,
\be\label{eq:Dirac_eqn_EM}
\gamma^\mu(\hbar\p_\mu-e a_\mu)\psi=0
,\quad
a_\mu=(v^{-1}\phi,a_1,a_2)
.
\ee
The potentials 
describe external fields $\vec E=-\nabla\phi$ and $B=\nabla\times\vec a$, pointing in-plane and out-of-plane, respectively, which can be either static or dynamical.
The analogy with $2+1$ Quantum Electrodynamics (QED), manifest in Eq.(\ref{eq:Dirac_eqn_EM}), helps to establish connections between various problems involving electrons in graphene with analogous problems in QED, described by the relativistic Dirac equation.

One aspect of the relation with QED which is particularly intriguing is the large value of the dimensionless coupling strength, the analog of the fundamental fine structure constant for electrons in graphene. Indeed, since $c/v\approx 300$, we have
\[
\alpha=\frac{e^2}{\hbar v\kappa}=\frac{c}{v}\times \frac{e^2}{\hbar c\,\kappa} = \frac{300}{137\kappa}\approx \frac{2.19}{\kappa}
\]
where the dielectric constant $\kappa$, describing the cumulative effect of screening by the substrate and intrinsic screening in graphene, typically can take values between 3 and 10 \cite{Fogler08a}. The large value of the fine structure constant, $\alpha\sim 1$, makes the QED-like physics in graphene even more rich and interesting than in ``natural'' QED systems.

As a cautionary remark we note that, while the 
analogy with $2+1$ QED is valid for coupling to arbitrary external fields, and is indeed useful in a variety of problems, it does not extend to the dynamical EM field caused by charges in graphene. The basic reason for this difference is the three-dimensional character of such EM fields, which originate on the charges in graphene, but spread out in the entire space outside the graphene plane. Hence the problem of electron interactions in graphene requires insights which are outside the realm of $2+1$ QED [e.g., see Refs.\cite{Son07,Gonzalez94,Biswas07} and references therein].

In this article we review some of the recent work on graphene motivated by analogy with QED, focusing on the phenomena familiar from relativitic quantum mechanics and on their connection to electronic transport.

In Sec. \ref{sec:collapse} we shall discuss the problem of atomic collapse, in which the large value of $\alpha$ is crucial. Atomic collapse is a phenomenon discovered theoretically in atomic physics, and predicted to occur in heavy atoms, having nuclear charge $Z\gtrsim 170$. Because such large values of $Z$ are not realized in any of the known elements, this phenomenon has not been experimentally observed. As we shall see, in graphene the condition for collapse, $Z\gtrsim \alpha^{-1}\sim 1$, is much easier to realize. In particular, collapse can occur for the electronic states near charge impurities, with critical charge as low as $Z_c=2$ \cite{Collapse1,Collapse2}.

In Sec. \ref{sec:Lorentz} we consider another system for which the connection with relativistic dynamics turns out to be fruitful: graphene p-n junctions. We show that transport in a p-n junction in the presence of a magnetic field can be understood with the help of a Lorentz transformation, which is a symmetry of the Dirac equation (\ref{eq:Dirac_eqn_EM}). This transformation, in which the velocity $v$ replaces the speed of light, can be used to eliminate either the magnetic field or the electric field, whichever is weaker. We compare this method to other approaches, and use it to find the field-dependent conductance of the system \cite{Lorentz}.

Next, in Sec. \ref{sec:FP}, we discuss transport in p-n-p structures. We focus on the ballistic regime, in which conductance exhibits Fabry-P\'erot (FP) resonances due to interference of electron waves reflected from two p-n boundaries. This system is convenient for investigating the Klein scattering phenomenon, which is a manifestation of chirality conservation in the dynamics of relativistic Dirac particles. In particular, because the reflection coefficient vanishes for normally incident particles, the backreflection amplitude, an analytic function of the incidence angle, changes sign at normal incidence. This sign change contributes a $\pi$ phase shift to the FP interference, which can be revealed by a half-a-period shift of FP fringes induced by a relatively weak magnetic field \cite{Klein_in_FP}. These predictions are in agreement with recent experiment \cite{Young08}.


\section{Atomic collapse and supercritical charge impurities}
\label{sec:collapse}

As we mentioned above, graphene offers an opportunity to observe the behavior of 
relativistic particles in very strong fields, a regime which is 
very difficult to achieve in high-energy experiments because 
of the small value of the fine structure constant. This is exemplified by
an interesting phenomenon, which will be discussed in this section, the collapse of a Dirac particle moving in a $1/r$ potential. 

To gain some intuition about this problem, let us start by recalling the explanation of  stability of the hydrogen atom in the framework of Bohr's quantum theory. Atomic stability results from quantum zero-point motion of an electron which prevents it from falling on the nucleus. Indeed, starting from the nonrelativistic Schr\"odinger equation, and estimating the kinetic energy of an electron as $E_{\rm kin}=\hbar^2/2m\rho^2$, where $\rho$ is the characteristic radius of electron wavefunction, we can write the total energy as
\be\label{K+V_nonrelativistic}
E(\rho)=\frac{\hbar^2}{2m\rho^2}-\frac{Ze^2}{\rho},
\ee
where $Z$ is nuclear charge. As a function of $\rho$, the energy (\ref{K+V_nonrelativistic}) 
is dominated by the positive kinetic energy term at small $\rho$, and by the negative Coulomb term at large $\rho$, giving rise to a minimum at $\rho_0=\hbar^2/(Ze^2m)$, the Bohr's radius. The latter quantity determines the size of an atom. 

The effects of quantum zero-point motion
which prevent an electron from falling onto the nucleus are less powerful
in the relativistic regime. 
In this case,
the balance between kinetic and potential energy becomes more delicate, because the kinetic energy $\epsilon= cp\sim c\hbar/\rho$ scales in the same way as the Coulomb energy. To see this more clearly, let us extend the reasoning used in (\ref{K+V_nonrelativistic}) to a relativistic electron, described by the kinetic energy $\epsilon(p)=\sqrt{(cp)^2+(mc^2)^2}$. Estimating $p\sim \hbar/\rho$, we again seak to minimize the sum of the kinetic and potential energies:
\be\label{K+V_relativistic}
E(\rho)=
\sqrt{\lp\frac{c\hbar}{\rho}\rp^2+(mc^2)^2}-\frac{Ze^2}{\rho}.
\ee
Looking for extrema of this function, we find an equation
$\sqrt{1+(mc/\hbar)^2\rho^2}=\hbar c/Ze^2$, which has a solution only when $Z<\hbar c/e^2\approx 137$. At higher values of $Z$ the effective potential (\ref{K+V_relativistic}) is a monotonic function varying between $E=mc^2$ at $\rho=\infty$ and $E=-\infty$ at $\rho=0$. Thus there is no stable solution for a hydrogenic atom at large nuclear charge.

The difference between the high and low values of $Z$ can be also seen directly in the exact solution of the Dirac equation for a hydrogenic atom,
\be\label{Dirac_eigenvalues}
E_{n,j}=mc^2\lb 1+\frac{(Z\alpha)^2}{\lp n-|j| +\sqrt{j^2-(Z\alpha)^2}\rp^2}\rb^{-1/2}
,
\ee
with the quantum numbers $n=1,2...$, $j=\pm1,\pm2...$, and $\alpha=e^2/\hbar c\approx 1/137$. The expression under the square root becomes negative when $Z\alpha>1$, rendering the eigenvalues (\ref{Dirac_eigenvalues}) with $j=\pm 1$ complex-valued. The unphysical complex energies indicate an intrinsic problem arising in the Dirac equation at supercritical $Z>\alpha^{-1}$.

Mathematically speaking, this behavior can be understood as an effect of the $1/r$ singularity of the Coulomb potential, making the Dirac operator non-Hermitian and leading to the breakdown of the Dirac equation. One could replace the $1/r$ potential by a regularized potential, rounded on a characteristic nuclear radius, which can be achieved by using a suitable nuclear form factor \cite{Smorodinsky}. However, it was found that, after the $1/r$ singularity is eliminated and the Dirac equation becomes well defined as a mathematical problem, the anomalous behavior at large $Z$ persists. 

\begin{figure}
\includegraphics[width=3.1in]{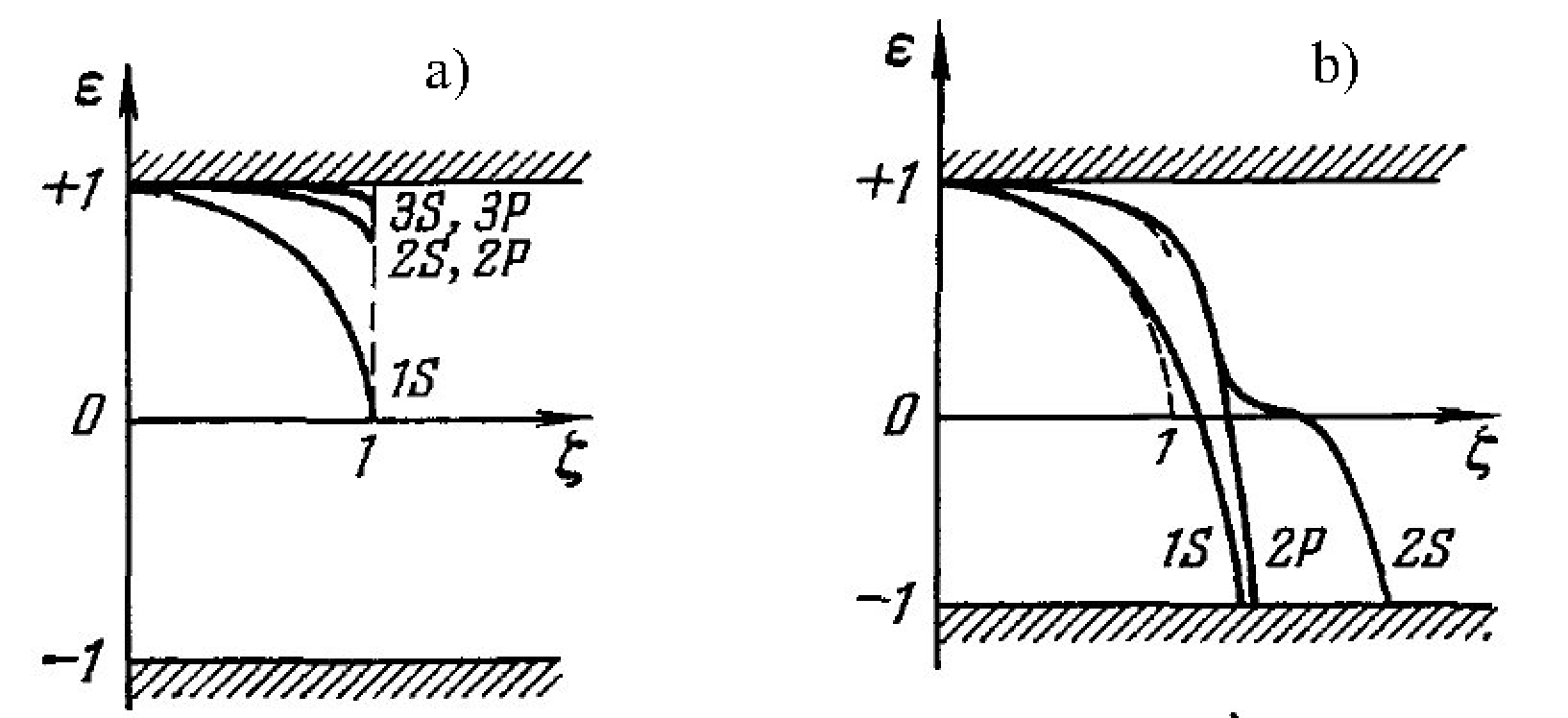} 
\caption[]{a) Energy levels of superheavy atoms obtained from Dirac equation for Coulomb potential $-Ze^2/r$, plotted as a function of $\zeta=Z\alpha$, where $Z$ is nuclear charge, and $\alpha=e^2/\hbar c $ is the fine structure constant.
Energy is in the units of $mc^2$. (b) Energy levels for Coulomb potential regularized on the nuclear radius. As $Z$ increases, the discrete levels approach the continuum of  negative-energy states and dive into it one by one at supercritical $Z>170$ (from Ref.\cite{Popov}).}
\label{fig2}
\end{figure}

This problem was analyzed half a century ago~\cite{Smorodinsky,Popov}, and it was predicted that in superheavy atoms with $Z > 170$ relativistic effects lead to a reconstruction
of the Dirac vacuum. 
As the nuclear charge $Z$ increases, it was found that the energies of discrete states approach the negative energy continuum, $\epsilon<-mc^2$, and then dive into it, one after another (see Fig.\ref{fig2}). After entering the continuum, discrete states turn into resonances with a finite lifetime, which can be described as resonant (or, quasistationary) states with complex energies. 

The finite lifetime of electronic states, which indicates an instability of superheavy atoms, puts a natural limit on the extent of the periodic table of elements. As such, the prediction of this phenomenon caused a lot of interest and excitement. Experiments on heavy-ion collisions probing this collapse 
have been suggested and attempted \cite{Popov01}; however, the results were
ambiguous and difficult to interpret. 

Graphene offers a completely new perspective on the problem of supercritical atoms, with charged impurities providing a natural realization of the Coulomb potential. 
Charge carriers in graphene, which are described by a \emph{massless} Dirac equation, mimic 
Dirac electrons in the ultrarelativistic regime, $\epsilon\gg mc^2$. 
In this case, 
an analysis of the balance between the zero-point motion and Coulomb attraction similar to that leading to Eq.(\ref{K+V_relativistic}) yields
%
\be\label{K+V_graphene}
E(\rho)=\frac{\hbar v}{\rho}-\frac{Ze^2}{\rho}=\frac{\hbar v-Ze^2}{\rho}.
\ee
%
For $Z>\alpha^{-1}=\hbar v/e^2$, the energy can be driven to arbitrarily large negative values by letting the radius $\rho$ tend to 0; thus we find a break-down or ``collapse'' of ultra-relativistic atoms in the supercritical regime $Z > \alpha^{-1}$.
Due to the low
value of the ``speed of light'' $v$, the critical charge
in graphene is of the order of one~\cite{Collapse1}. This makes charged impurities in graphene a very convenient system for experimental investigation of this phenomenon.

Electronic states near a charged impurity are described by 
the two-dimensional Dirac-Kepler problem
\be\label{Dirac-Kepler}
 \hbar v \lp \begin{array}{cc}
0 & -i\p_x-\p_y\cr -i\p_x+\p_y& 0
\end{array}\rp \psi = \lb \epsilon-\frac{Ze^2}{\kappa\rho}\rb \psi.
\ee
%
Properties of the solutions of Eq.(\ref{Dirac-Kepler}) depend on the dimensionless parameter $\beta=Ze^2/\hbar v\kappa$, exhibiting two different regimes. For $|\beta|<1/2$, the behavior of states 
is consistent with scattering on a weak potential. For $|\beta|>1/2$, there is an abrupt reconstruction of the Dirac vacuum around the $1/r$ potential, which leads to the formation of resonant states lying in the continuum. 
These states can be understood as discrete electronic states, broadened into resonances by their coupling to the continuum of hole states \cite{Collapse2}. 
Comparing this behavior to the predictions for superheavy atoms \cite{Smorodinsky,Popov}, we can identify these two regimes with
\be\label{two_regimes}
{\rm no\ collapse}\ (|\beta|<1/2)
\ {\rm and\ collapse}\ (|\beta|>1/2).
\ee
Note the critical value $1/2$ instead of $1$ found in three dimensions. 

The formation of resonances 
can be understood using scattering states of the Dirac-Kepler problem. Using polar coordinates, $x+iy=\rho e^{i\phi}$, the general solution of Eq.(\ref{Dirac-Kepler})
can be written as a sum of cylindrical waves with different angular quantum numbers \cite{Collapse1} 
\be\label{scattering states}
\psi(\rho,\phi)=\lp\begin{array}{c} 
w(\rho)+v(\rho) \cr [w(\rho)-v(\rho)] e^{i\phi}
\end{array}\rp
\rho^{s-1/2}e^{i(m-1/2)\phi} e^{ik\rho}
\ee
where $s=\sqrt{m^2-\beta^2}$, $k=-\epsilon/\hbar v$, and $m$ is a half-integer angular quantum number. Here $w$ and $v$ are the amplitudes of the incoming and outgoing cylindrical waves, satisfying a hypergeometric equation. Solutions of this equation, combined with a boundary condition at short distances, can be used to construct scattering states. A ``zigzag'' boundary condition was used in \cite{Collapse1}, with $\psi_2(\rho)=0$ at the radius $\rho = r_0$ on the order of carbon atom spacing.
Other boundary conditions yield similar behavior.

\begin{figure}
\includegraphics[width=3.1in]{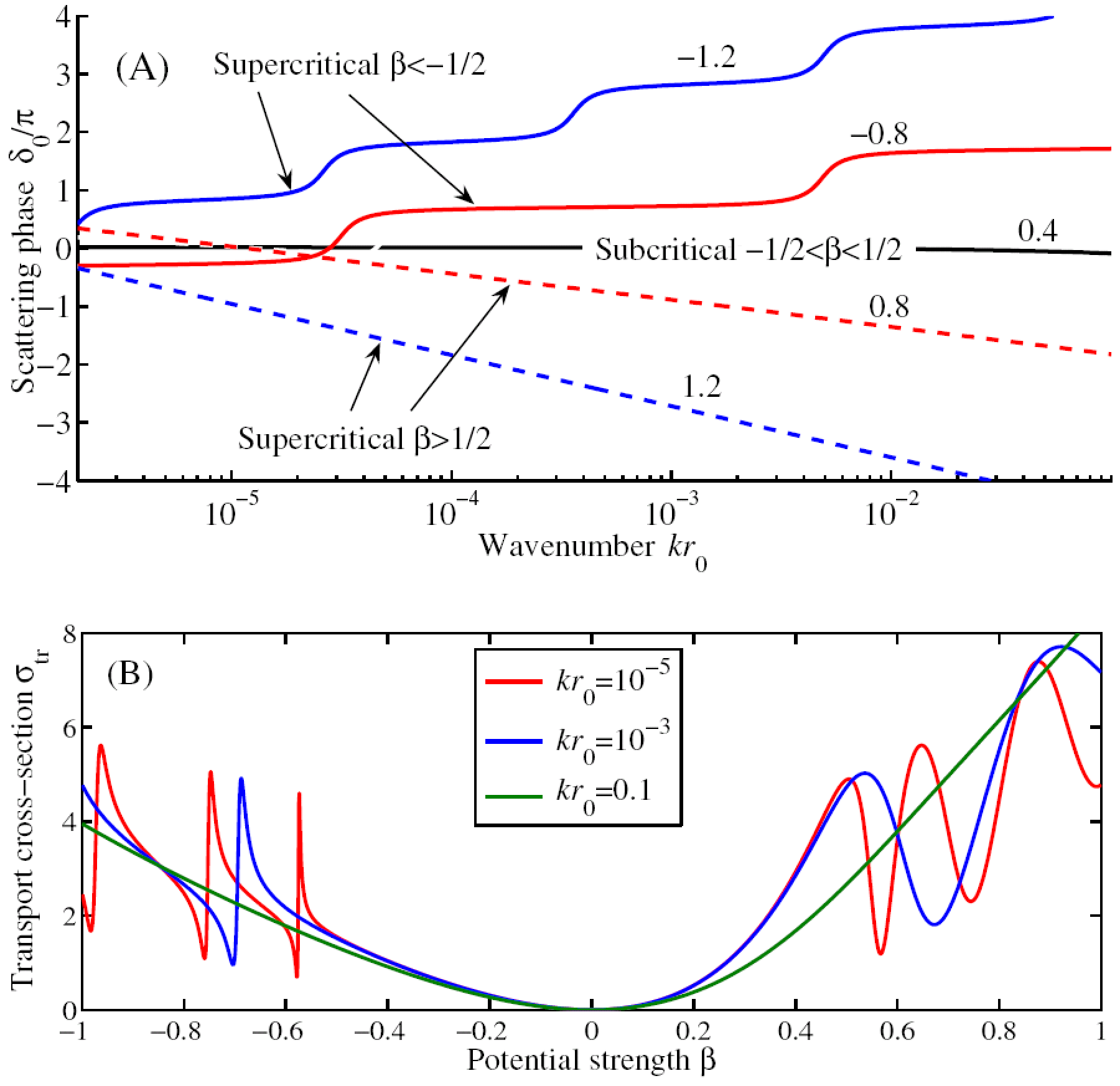} 
\caption[]{(A) The dependence of the scattering phases $\delta_m$ on energy $\epsilon=-\hbar v/k$, shown for several subcritical and supercritical values of $\beta=Ze^2/\hbar v$ (from Ref.\cite{Collapse1}). For subcritical $\beta$, the phases $\delta_m$ have no energy dependence (black line). For supercritical $\beta$, the phases acquire energy dependence, which becomes stronger as $|\beta|$ increases. For $\beta<-\frac12$ (Coulomb attraction), the kinks in $\delta_m(k)$, indicate formation of resonant states with complex energies, ${\rm Re}\,\epsilon<0$, corresponding to Breit-Wigner resonances in scattering. For $\beta>\frac12$ (Coulomb repulsion), the dependence $\delta_m$ vs. $\log k$ is approximately linear, exhibiting no kinks. (B) The transport cross-section, Eq.(\ref{eq:sigma_transport}), shown as a function of $\beta$ for several energy values. Note the Fano-shaped resonances at negative $\beta<-\frac12$, appearing due to resonant states (kinks in (A)), and oscillations at $\beta>\frac12$, resulting from the energy-dependent $\delta_m$.}
\label{fig3}
\end{figure}

Scattering phases $\delta_m(k)$, obtained from the asymptotic ratio $v/w=e^{2i\delta_m(k)+2ik\rho}$ ($k\rho\gg1$), exhibit very different behavior in the two regimes (\ref{two_regimes}), illustrated in Fig.\ref{fig3}. The resonant states with complex energies, formed at $\beta<-\frac12$, appear as kinks in the dependence $\delta_m$ vs. $k$. 

The resonant states should manifest themselves experimentally via various transport properties. In particular, the transport scattering cross-section
\be\label{eq:sigma_transport}
\sigma_{\rm tr}=\frac{4}{k}\sum_{m=0}^\infty \sin^2(\delta_{m+1}-\delta_m)
,
\ee
exhibits a resonant structure which is shown in Fig.\ref{fig3}B. In contrast, in the subcritical regime $|\beta|<\frac12$ the differences $\delta_{m+1}-\delta_m$ are energy-independent, which makes the cross-section scale as $1/k$ with a slight asymmetry in the prefactor for positive and negative $\beta$ values, which was discussed in \cite{Novikov07}. 

The most direct signature of resonant states, however, can be seen in the local density of states (LDOS), 
\be\label{DOS}
\nu(\epsilon,\rho)=\frac4{\pi\hbar v}\sum_m|\psi(k,\rho)|^2
\ee
where $\psi(k,\rho)$ are the scattering states (\ref{scattering states}). This quantity can be directly measured by energy-resolved scanning tunneling spectroscopy probes.

\begin{figure}
\includegraphics[width=2.5in]{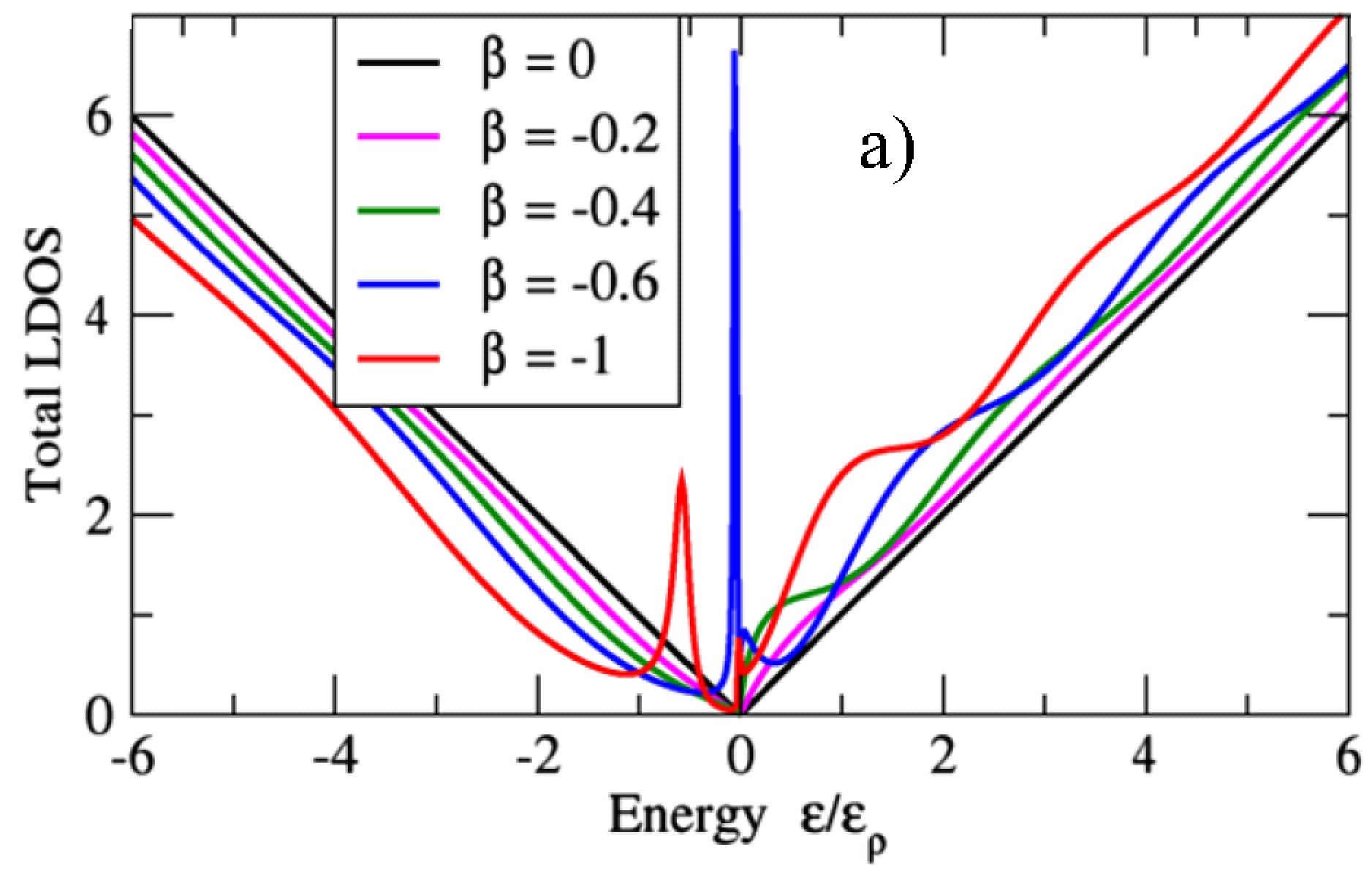} 
\includegraphics[width=3.6in]{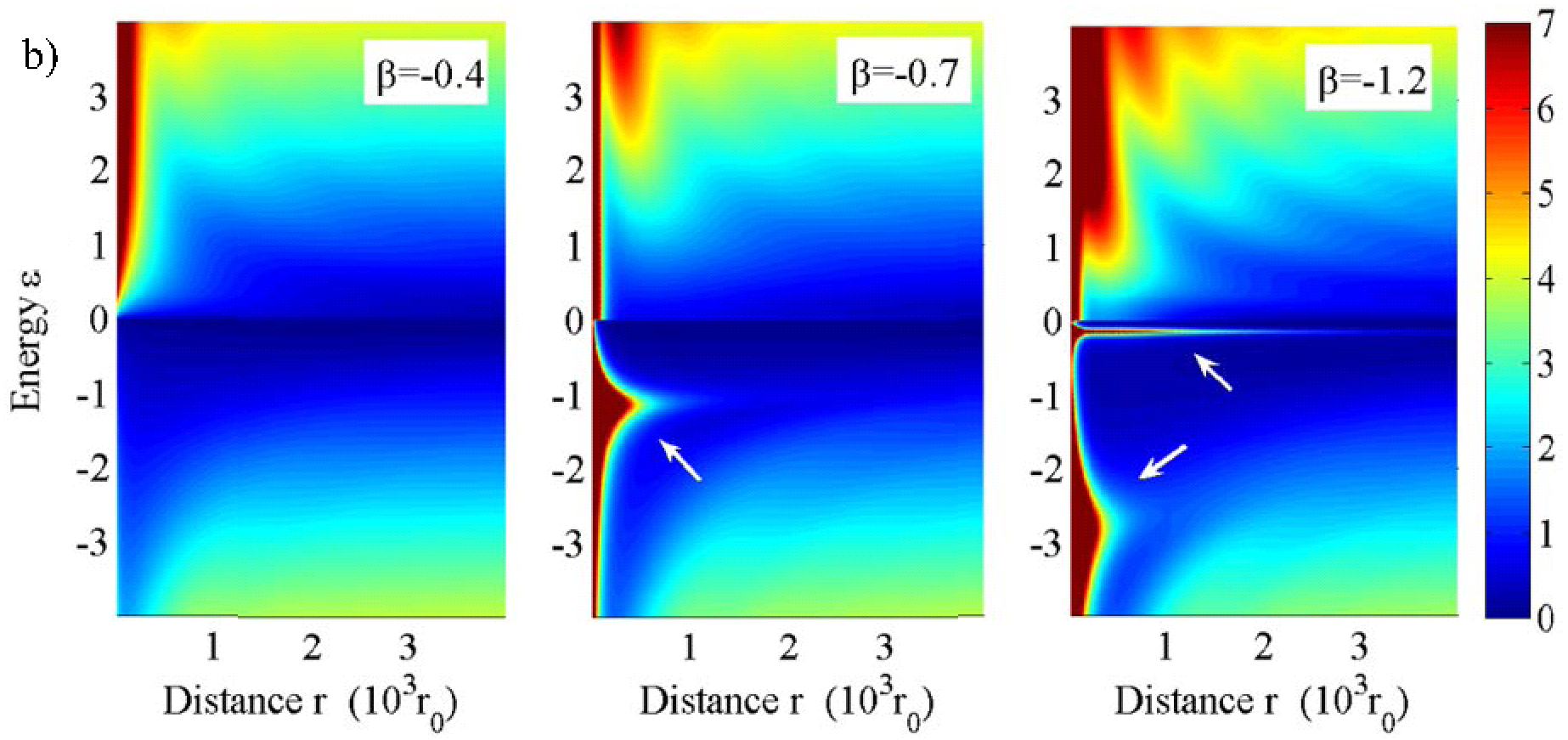} 
\caption[]{(a) Local density of states (\ref{DOS}) calculated at a fixed distance $\rho=10^3r_0$ from the charged impurity, where $r_0$ is a short-distance parameter of the order of carbon lattice spacing (from Ref.\cite{Collapse1}). Peaks in the LDOS, which appear at supercritical $\beta$ and move to more negative energies at increasing $|\beta|$, correspond to the resonant states. (b) Spatial map of the density of states, shown for several values of $\beta$, with resonances marked by white arrows (from Ref.\cite{Collapse2}). Note that the spatial width of the resonances decreases at they move to lower energies, 
$\Delta\rho\propto 1/|\epsilon|$, while the linewidth increases, $\gamma\propto |\epsilon|$. The oscillatory structure at positive energies represents standing waves with maxima at $k\rho\approx (n+\frac12)\pi$, similar to those studied in carbon nanotubes \cite{Ouyang02}. Energy is given in the units of $\epsilon_0=10^{-3}\hbar v/r_0\approx 30\,{\rm mV}$ for $r_0=0.2\,{\rm nm}$. 
}
\label{fig4}
\end{figure}

The energy dependence, Eq.(\ref{DOS}), calculated at a fixed distance from the charged impurity, is illustrated in Fig.\ref{fig4}. Overall, the LDOS is an approximately linear function of energy, with oscillations at positive energies and sharp resonances at negative energies, which appear at supercritical $\beta$. The resonances, which occupy a spatial region of size $\rho \approx \hbar v/|\epsilon|$ near a charged impurity, provide a clear signature of atomic collapse accessible with scanning probe techniques.

To estimate the critical value of impurity charge $Z_c=1/2\alpha$, we need to find $\alpha=e^2/\hbar v\kappa$ taking into account screening. Screening due to the dielectric substrate with dielectric constant $\epsilon$ is described by $\kappa=(\epsilon+1)/2$. More interesting, however, is the intrinsic screening due to polarization of the graphene Fermi sea. Using random-phase approximation (RPA) for the dielectric function of undoped graphene, $\kappa_{\rm RPA}\approx 5$, and ignoring the effect of dielectric substrate, we find that $Z=1$ is subcritical, while $Z=2$ is supercritical \cite{Collapse1}. This conclusion is in agreement with recent work \cite{Terekhov08} which analyzes screening of charged impurity as polarization of Dirac vacuum in a strong Coulomb field (see also Refs. \cite{Biswas07,Kotov08}). 

It was noted in Ref. \cite{Collapse1} that linear screening breaks down in the supercritical case, for undoped graphene. The reason is that the vacuum polarization in the presence of supercritcal $1/r$ potential, found from the solution of the noninteracting problem which we considered above, has a $1/r^2$ profile. This leads to a log divergence of the polarization charge, and thus to overscreening, unless the effects of interaction within the polarization cloud are taken into account. It was shown with the help of a renormalization group argument that the true value of polarization charge is $-(Z-Z_c)$, i.e. the polarization cloud compensates the excess part of the impurity charge and brings it down to the critical value. 

The profile of the induced polarization can be found approximately for $e^2/\hbar v\ll1$ \cite{Collapse1}, and also for $e^2/\hbar v\gg1$ \cite{Fogler07}. The effective potential, resulting from adding a contribution of polarization charge to the bare $1/r$ potential, deviates from Coulomb potential, and thus the above solution of the Dirac-Kepler problem strictly speaking is not applicable. Still, the qualitative picture described above, with resonant states appearing in the Dirac continuum at $Z>Z_c$ remains valid even in the nonlinear screening regime.








\section{Lorentz transformation and collimated transmission in p-n junctions.}
\label{sec:Lorentz}

A basic property of massless Dirac particles is conservation
of chirality, defined as the projection of pseudospin on the velocity vector. 
Conserved chirality leads to unimpeded tunneling through 
arbitrarily high barriers (the so-called Klein phenomenon) \cite{Klein29}; particles with conserved chirality, incident normally on a barrier, exhibit perfect transmission and zero reflection \cite{Katsnelson06,Cheianov06}.
The Klein effect allows electrons to leak out of any trap \cite{Silvestrov07,Pereira07b}, 
and makes electrostatic confinement in graphene more difficult than, 
e.g., in gated semiconductor systems. 
Klein transmission is also an essential part of theoretical understanding of transport in systems such as graphene p-n junctions \cite{Beenakker07}. 

The range of angles in which transmission can occur depends on the microscopic details of a potential barrier \cite{Katsnelson06}. For smooth barriers, such as those in which the potential varies on a scale large compared to the Fermi wavelength, perfect transmission is restricted to a narrow range of angles. This is the case for p-n junctions created by electrostatic gates, which are typically placed few tens of nanometers above graphene plane \cite{Huard07,Williams07,Ozyilmaz07}.

Collimated transmission in such systems can be analyzed by replacing the externally imposed electrostatic field in the region where Klein tunneling occurs by a uniform in-plane electric field \cite{Cheianov06}. 
For massless Dirac particles moving in a uniform in-plane electric field $E\parallel \vec x$, the transmission coefficient, as a function of electron momentum projection on the p-n interface, is given by
\be\label{eq:transmission_B=0}
t(p_y)=e^{-\lambda p_y^2}
,\quad
\lambda=\pi v/\hbar |eE|
\ee
The range of $p_y$ in which transmission can occur becomes very narrow for small $E$, i.e. for smooth potentials.

What happens to the collimated Klein transmission (\ref{eq:transmission_B=0}) in the presence of a magnetic field? The Dirac equation (\ref{eq:Dirac_eqn_EM}), describing an electron in graphene in the presence of external fields, exhibits
two qualitatively different regimes, depending on the relative strength of the $E$ and $B$ fields. For weak magnetic fields, $|B|<|E|/v$, the picture of collimated transmission described in Ref.\cite{Cheianov06} remains essentially unchanged, with perfect transmission occurring for the incidence angle which is a function of magnetic field, 
%
\be\label{eq:arcsin(B/E)}
\theta_0=\arcsin(vB/E)
,\quad t(\theta_0)=1
.
\ee
In contrast, at stronger magnetic field, $v|B|>|E|$, electronic states are described by quantized Landau levels with position-dependent energies \cite{Lukose07}. In this regime, in a clean system, transport can occur only perpendicular to the electric field.

There are several ways to understand the origin of the two regimes. Perhaps
most easily and elegantly, it can be seen by employing Lorentz invariance of the equations of motion, Eq.(\ref{eq:Dirac_eqn_EM}), with respect to Lorentz transformations in which the graphene dispersion velocity $v$ plays the role of the speed of light. There is a basic fact, established in special relativity, that for perpendicular electric and magnetic fields, a Lorentz boost can be performed such that in the moving frame one of the fields vanishes, while the other field remains nonzero, experiencing Lorentz contraction. 
The field which can be eliminated by such Lorentz boost is the weaker of the two. Thus, for $v|B|<|E|$, one can eliminate the magnetic field by performing a boost to a moving frame with velocity $u=v^2B/E$. For $v|B|>|E|$, the electric field can be eliminated by a boost with velocity $u=E/B$.
In the first case, with $B$ zero and  $E$ nonzero in the moving frame, we have a situation identical to that studied in Ref.\cite{Cheianov06}. 
In the second case, with $E=0$ in the moving frame, and $B$ finite, electronic states are derived from quantized Landau levels. 

It may be useful to point out that the situation with nonrelativistic, Galilean transformations is quite different. In the presence of perpendicular electric and magnetic fields, a Galilean transformation can be used to eliminate $E$ by performing a boost in the moving frame with velocity $u=E/B$. It is not possible, however, to eliminate magnetic field by a Galilean transformation.

Given this property of Galilean transformations, one may be suspicious about drawing conclusions from pseudo-Lorentz transformations, which appear to be a kind of mathematical trick, rather than a genuine symmetry of the system. To relieve this concern, we now present another argument leading to the two regimes, which does not rely on Lorentz invariance (following a suggestion by F. D. M. Haldane).

Let us consider classical trajectories of a charged particle with kinetic energy $\epsilon(\vec p)$, which is moving in the fields $E$ and $B$. This problem can be described by canonical equations of motion with a Hamiltonian
\[
\HH(p,r) = \epsilon(\vec p) -eEx
,\quad
\vec p = \tilde{\vec p} -e\vec A
,\quad
\vec A=(0,Bx)
,
\]
where we have chosen the gauge so that the vector potential is parallel to the vector ${\vec v}_{\rm d} = \vec E\times \vec B/B^2$, which defines classical drift velocity. With this choice of $\vec A$, and taking into account conservation of the component of momentum $\tilde{ p}_2$ perpendicular to ${\vec v}_{\rm d}$, it is straightforward to write an equation for particle trajectory in momentum space, given by the energy integral
\[
\epsilon(\vec p)-{\vec v}_{\rm d} . \vec p=\epsilon_0
,
\]
where we used the relation $p_2=\tilde p_2-eBx$ to express $x$ through $p_2$, absorbing constant $\tilde p_2$ in $\epsilon_0$.
In the case of massless Dirac particles, with the kinetic energy $\epsilon(\vec p)=\pm v|\vec p|$, we can write the trajectory in polar coordinates in momentum space as
\[
p(\theta)=\frac{\epsilon_0}{\pm v -v_{\rm d}\cos\theta}
\]
where $\theta$ is the angle measured relative to $\vec v_{\rm d}$.
This formula is nothing else than an equation for conical sections, familiar from the theory of planetary motion (an electron is orbiting the Dirac point in a manner similar to a planet, or a comet, orbiting the Sun). Depending on the relative magnitude of $v$ and $v_{\rm d}$, there are two cases, 
\[
v_{\rm d}=E/B>v\ {\rm (hyperbola)},\quad v_{\rm d}=E/B<v\ {\rm (ellipse)},
\]
which coincide with the two regimes identified using Lorentz transformations.
The trajectories are open for $v_{\rm d}>v$, and closed for $v_{\rm d}<v$, corresponding, respectively, to motion that originates at infinity and ends at infinity, and to periodic cyclotron motion. 

Now we proceed to analyze transmission in the regime $v|B|<|E|$. As discussed above, we can eliminate $B$ by 
a Lorentz boost with rapidity $\beta = -vB/E$ parallel to $\vec v_{\rm d}$:
\be\label{eq:L-boost}
\Lambda=\left(
\begin{array}{ccc}
\gamma & 0 & \gamma\beta  \\
0 & 1 & 0 \\
\gamma\beta & 0 & \gamma 
\end{array}
\right)
,\quad \gamma=\frac1{\sqrt{1-\beta^2}}
.
\ee
In the new frame we have $B'=0$,
$E'=E/\gamma$.

Since $B'=0$, we can use Eq.(\ref{eq:transmission_B=0}), with $E$ replaced by $E'$, giving
the transmission coefficient as a function of momentum $p'_y$
parallel to $\vec v_{\rm d}$. Expressing $p'_y$ and $E'$ through the quantities in the lab frame, we obtain
%
\be\label{eq:T(k,epsilon)}
T(p_y)= e^{ -\lambda \gamma^3 (p_y+\beta\tilde\epsilon)^2}
,\quad
\tilde\epsilon=\epsilon/v
.
\ee
Introducing the incidence angle, $vp_y=\epsilon\sin\theta$, we find the value $\theta=\arcsin(vB/E)$ for which the transmission is perfect, in agreement with Eq.(\ref{eq:arcsin(B/E)}). The angular size of transmitted beam is a function of $B$, such that the beam becomes more collimated at higher $B$.

In passing between the moving and lab frames we used the fact
that the transmission coefficient, 
Eq.(\ref{eq:T(k,epsilon)}), 
is a scalar with respect to Lorentz transformations, Eq.(\ref{eq:L-boost}).
This is indeed true because transmission and reflection 
at the p-n interface
is interpreted in the same way
by all observers moving 
with velocity $\vec u\parallel \vec v_{\rm d}$. 

The conductance can be found by integrating the transmission (\ref{eq:T(k,epsilon)}) over $p_y$, which gives \cite{Lorentz}
\be\label{eq:G(B)}
G(B)=\cases{G_0\lp 1-B^2/B_0^2\rp^{3/4} & $B<B_0$\cr 0 & $B>B_0$}
,
\ee
where $B_0=|E|/v=\pi/\hbar |e|\lambda$ and $G_0$ is the conductance of a ballistic p-n junction at $B=0$, proportional to the length of the p-n interface, which was found in \cite{Cheianov06}. 

The critical field value $B_0$ can be estimated from density profile in the p-n junction region, which can be found from electrostatic simulation, as discussed in Refs.\cite{Fogler08a,Gorbachev08}. Using the relation $n=k_F^2/\pi$, we estimate the electric field in the junction as $eE\approx 2\hbar v k_F/d=2\hbar v\sqrt{\pi n}/d$, where $d$ is the width of the density step. This gives
$B_0\approx  \Phi_0\sqrt{n/\pi}/d$, where $\Phi_0=h/e$ is the flux quantum.
For $n=10^{12}\,{\rm cm^{-2}}$ and $d=100\,{\rm nm}$ we estimate $B_0\approx 2.3\,{\rm T}$.

Caution should be used when the result (\ref{eq:G(B)}), describing the field dependence of conductance in the ballistic regime,
is applied to real systems, such as those studied in Refs.\cite{Gorbachev08,Stander08}. First, our derivation only accounts for the current flowing through the entire cross-section of the system, and not for, e.g., the edge currents. The latter become important at strong fields, $B>B_0$, when the system enters the quantized Hall regime, giving a contribution to conductance of the order of $e^2/h$. 

In addition, in studying a p-n junction in the presence of disorder, one should account for net contribution of disordered regions to total resistance. A series resistance model was proposed in Ref.\cite{Fogler08b} to describe this situation. However, the magnetic field dependence in the disorder-dominated transport regime has not yet been analyzed.

\section{Klein scattering and Fabry-P\'erot resonances in p-n-p structures}
\label{sec:FP}

Theoretically, Klein scattering is regarded as a fundamental mechanism of transport through potential barriers and p-n junctions in graphene \cite{Katsnelson06,Cheianov06}. Until recently, however, this theoretical picture was disconnected from experiment.  The manifestations of Klein scattering that have been discussed so far, such as collimated transmission  
which becomes perfect at normal incidence, can be easily masked by scattering on disorder. Because disorder is an integral part of any realistic system, these predictions may be difficult to verify using existing transport data.
 
Recently, however, another manifestation of the Klein phenomenon in transport was discussed, which may offer a more direct experimental signature 
\cite{Klein_in_FP}. The idea is to refocus attention from Klein transmission to Klein backreflection, which must vanish at normal incidence. Since the reflection amplitude is a smooth function of incidence angle $\theta$, which vanishes at $\theta=0$, it must change sign when $\theta$ varies from positive to negative values. Furthermore, because the sign change translates into a $\pi$ phase shift of the reflection phase, it can be revealed by Fabry-P\'erot interference in a p-n-p structure.

\begin{figure}
\includegraphics[width=3.2in]{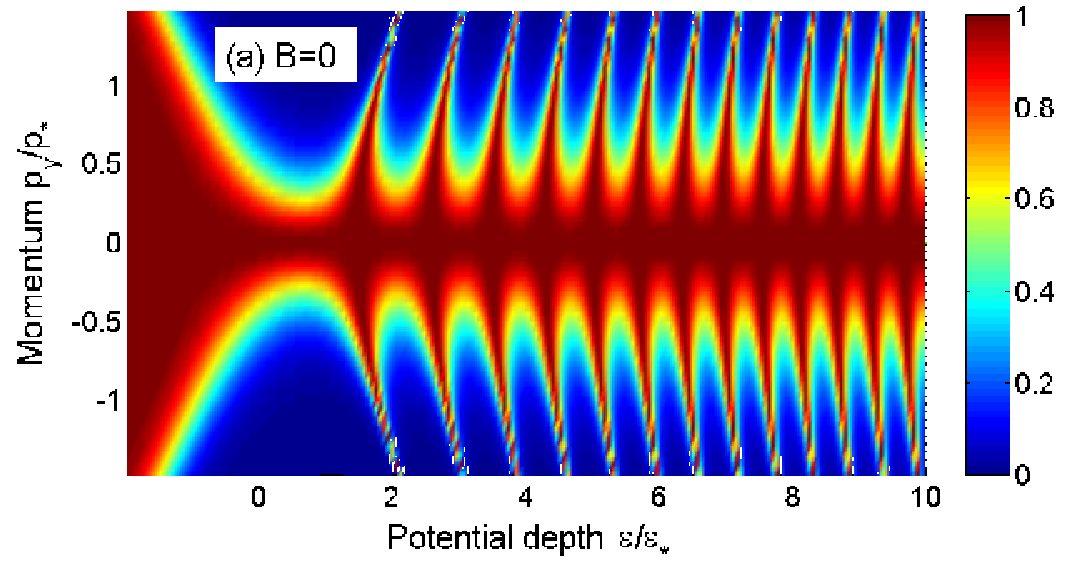} 
\includegraphics[width=3.2in]{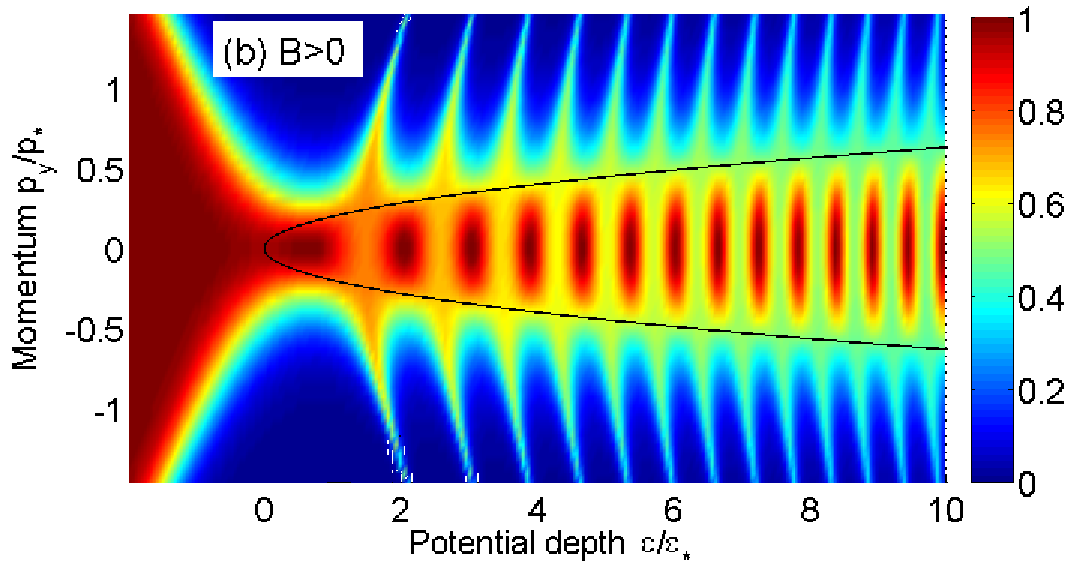} 
\caption[]{Transmission coefficient of p-n-p structure, obtained from numerical solution of the Dirac equation with potential $U(x)=ax^2-\epsilon$, plotted as a function of the component of electron momentum parallel to the p-n boundaries $p_y$ and potential depth $\epsilon$. At zero magnetic field (a), transmission exhibits fringes with a phase which is nearly independent of $p_y$. At finite magnetic field (b), fringe contrast reverses its sign on the parabola (black line), which marks the boundaries of the interval (\ref{minus_sign_range}) (from Ref.\cite{Klein_in_FP}). Here $\epsilon_*=(a\hbar^2 v^2)^{1/3}$, $p_*=\epsilon_*/v$.
}
\label{fig6}
\end{figure}

The simplest regime to analyze
is ballistic transport through a p-n-p structure. Transmission through two parallel p-n interfaces, which we label 1 and 2, 
can be described by the Fabry-P\'erot model,
\be\label{eq:FPgeneral}
T(p_{y}) = \frac{t_1t_2}{\left|1-\sqrt{r_1r_2}e^{i\Delta\theta}\right|^2}
\ee
where $t_{1(2)}$ and $r_{1(2)}=1-t_{1(2)}$ are the transmission and reflection coefficients for each interface.
The phase $\Delta\theta$ is a sum of the WKB phase and the phases of the reflection amplitudes,  
\be\label{eq:theta_total}
\Delta\theta = \frac2{\hbar}\int_1^2p_x(x')dx'
+\Delta\theta_1+\Delta\theta_2
.
\ee
%
Here $\Delta\theta_{1(2)}$ are the backreflection phases for the interfaces 1 and 2,
exhibiting a $\pi$-jump at zero incidence angle.

One could argue that, since the incidence angles for two parallel p-n interfaces are the same, the two $\pi$-jumps cancel each other in the net phase $\Delta\theta_1+\Delta\theta_2$, making the interference in (\ref{eq:FPgeneral}) insensitive to this effect. However, as pointed out in Ref.\cite{Klein_in_FP}, this cancellation can be eliminated by applying a magnetic field, which curves electron trajectories and makes the incidence angles unequal.

\begin{figure}
\includegraphics[width=3.0in]{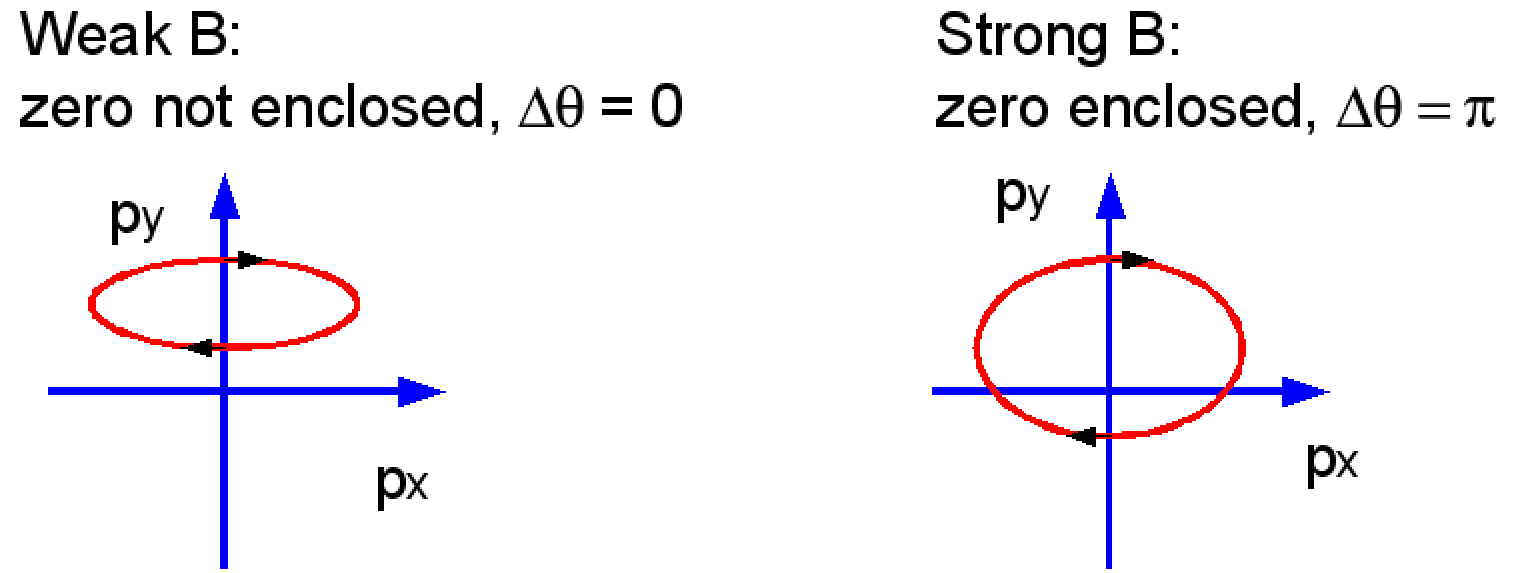} 
\caption[]{Berry phase for periodic orbit of an electron bouncing between p-n boundaries. As $B$ increases, with $\tilde p_y$ kept constant, the trajectory evolves from a skinny oval to a less skinny oval. For typical $\tilde p_y$ the trajectory does not enclose the origin at small $B$, giving zero Berry phase. At higher $B$, when the origin is enclosed, the Berry phase equals $\pi$.}
\label{fig5}
\end{figure}

The effect of the magnetic field can be taken into account by using the vector potential $\vec A=(0,Bx)$.  In this gauge, the $y$-component of electron canonical momentum (i.e. the component parallel to the p-n interfaces)
is conserved, and hence one can write the $y$-component of electron kinetic momentum as 
\be\label{eq:p kinetic-canonical}
p_y(x)=\tilde p_y - eBx
,
\ee
where the canonical momentum $\tilde p_y$ is a constant of motion that labels different scattering states. Using this relation, and evaluating the transmission coefficients on each interface $x=x_{1(2)}$ with the help of Eq.(\ref{eq:transmission_B=0}), we find
$t_{1(2)}=e^{-\lambda p^2_y(x_{1(2)})}$. From this, we obtain reflection amplitudes
\be\label{eq:WKB_r1}
{\rm sgn}\lb p_y(x_{1(2)}) \rb\,e^{i\theta_{\rm reg}(p_y)}\,\sqrt{1-e^{-\lambda p^2_y(x_{1(2)})}} 
,
\ee
where we factored out the sign, responsible for the phase jump,
and a regular part of the phase $e^{i\theta_{\rm reg}}$, as follows from analyticity in $p_y$. 

In the model used in Ref.\cite{Klein_in_FP}, the gate potential was described by a parabola $U(x)=ax^2-\epsilon$. In this case, $\epsilon>0$ creates p-n interfaces at
$x_{1(2)}=\pm\sqrt{\epsilon/a}$. Linearizing $U(x)$ near $x_{1(2)}$, we find the parameter $\lambda=\frac{\pi}2(a\epsilon)^{-1/2}$. 

The sign jumps in expressions (\ref{eq:WKB_r1}) occur when $p_y(x_{1(2)})=0$, where semiclassical electron trajectories approach the boundaries 1 and 2 at normal incidence. 
We can thus identify a range of values
\be\label{minus_sign_range}
eBx_1<\tilde p_{y}<eBx_2
\ee
for which the reflection amplitudes have opposite signs. For these values of $\tilde p_{y}$, the FP interference has a $\pi$ phase shift compared to what could be inferred from the WKB phase.

\begin{figure}
\includegraphics[width=3.6in]{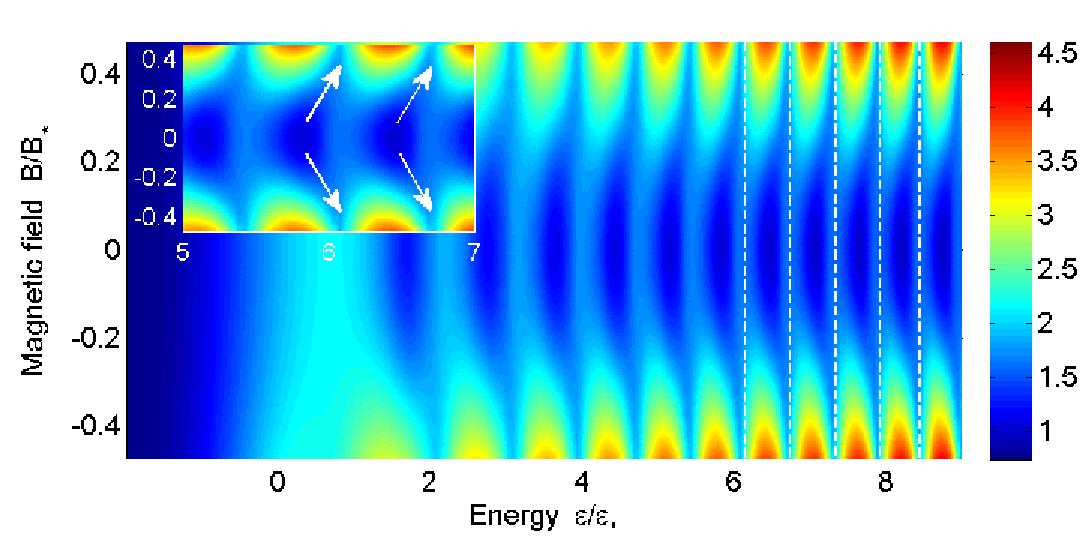} 
\caption[]{Resistance of a p-n-p structure, obtained from transmission, shown in Fig.\ref{fig6}. FP interference fringes in resistance as a function of potential depth $\epsilon$ exhibit half-a-period shift when magnetic field is applied. The units for magnetic field are $B_*=(\hbar/e)(a/\hbar v)^{2/3}$  
(from Ref.\cite{Klein_in_FP}).}
\label{fig7}
\end{figure}

Additional insight into the origin of the $\pi$ phase shift induced by magnetic field in FP interference is provided by a Berry phase argument, which focuses on the properties of quasiclassical trajectories. There is an analogy between the Dirac Hamiltonian (\ref{H_D}) and the Zeeman Hamiltonian of a spin $1/2$ in a time varying magnetic field, with the pseudospin, associated with $A$ and $B$ sublattices, playing the role of spin. If particle motion is periodic, over each period the wave function gains a phase given by a half of the solid angle swept by the effective magnetic field $\vec b_{\rm eff}(t)=v\vec p(t)$. In the case of cyclotron motion, this phase, combined together with the WKB phase, was used to explain half-integer quantization of Landau levels \cite{Novoselov05,Zhang05}.


Here we apply a similar argument to the trajectory of an electron bouncing between p-n boundaries. It is instructive to do the analysis
in momentum space, focusing on the evolution
with increasing $B$. At small $B$, since $p_y$ is nearly constant (see Eq.(\ref{eq:p kinetic-canonical})), the trajectory maps out a skinny oval, shown in Fig.\ref{fig5}. For typical values of $\tilde p_y$ this oval does not enclose the origin, and thus the Berry phase associated with it is zero. For larger $B$ the oval height increases, until eventually it encloses the origin. At this point the Berry's phase becomes equal to $\pi$.

This behavior can also be clearly seen in Fig.\ref{fig6}, in which the transmission coefficient, obtained from numerical solution of the Dirac equation with potential $U(x)=ax^2-\epsilon$, is displayed. Transmission exhibits FP interference fringes in the $\epsilon$ direction both at zero field (panel a) and at finite field (panel b). In the latter case the fringe contrast also exhibits sign reversal across the black line. This line is the parabola $p_y=\pm eB\sqrt{\epsilon/a}$, which marks the boundaries of the interval (\ref{minus_sign_range}), and separates regions with opposite phase contrast.

Contrast reversal directly manifests itself in conductance, which can be found by integrating transmission over $p_y$. Since the fringes in transmission are nearly vertical (Fig.\ref{fig6}), the integral over $p_y$ yields an oscillatory function of $\epsilon$. As the $B$ field increases, making the region with inverted contrast wider, the oscillations in conductance shift by approximately one half-period (see Fig.\ref{fig7}). The shift occurs in relatively weak fields $B\sim B_*=(\hbar/e)(a/\hbar v)^{2/3}$, which gives a few hundred of milliTesla for realistic parameter values.

Fabry-P\'erot oscillations in a ballistic p-n-p structure were reported in recent experiment \cite{Young08}. The behavior of the FP fringe contrast under applied magnetic field is reminiscent of that in Fig.\ref{fig7}, with a shift of approximately half-a-period observed in the fields of about $0.2-0.5\,{\rm T}$. The data \cite{Young08}  also exhibits an interesting crossover between FP resonances at $B=0$ to Shubnikov-deHaas oscillations at high $B$, which indicates that momentum-conserving transport through the structure, which is responsible for FP resonances, gives way to disorder-dominated transport at high magnetic fields.

In summary, graphene is a fertile system in which many interesting quantum-relativistic phenomena, or their condensed matter analogs, can be explored. As a material amenable to various experimental techniques, graphene offers new interesting connections between condensed matter and high-energy physics.

We thank F. D. M. Haldane, P. Kim,  A. K. Savchenko, and A. Young for useful discussions.

\end{document}